\documentclass[aps,prl,showpacs,amsmath,amsfonts,superscriptaddress,twocolumn]{revtex4}

\usepackage{graphicx}% Include figure files
\usepackage{dcolumn}% Align table columns on decimal point
\usepackage{bm}% bold math
\usepackage{color}

\def\={\, =\, }

\begin{document}

\author{Nikolay V. Prokof'ev}
\affiliation{Department of Physics, University of Massachusetts,
Amherst, MA 01003, USA}
\affiliation{Russian Research Center ``Kurchatov Institute'',
123182 Moscow, Russia}
\author{Boris V. Svistunov}
\affiliation{Department of Physics, University of Massachusetts,
Amherst, MA 01003, USA}
\affiliation{Russian Research Center ``Kurchatov Institute'',
123182 Moscow, Russia}

\title{From Popov-Fedotov trick to universal fermionization}

\begin{abstract}
We show that Popov-Fedotov trick of mapping spin-1/2 lattice systems on
two-component fermions with imaginary chemical potential readily generalizes to bosons with a fixed (but not limited) maximal site occupation number, as well as to fermionic Hamiltonians with various constraints on the site Fock states.  In a general case, the  mapping---fermionization---is on multi-component fermions with many-body non-Hermitian interactions. Additionally, the fermionization approach allows one to convert large many-body couplings into single-particle energies, rendering the diagrammatic series free of large expansion parameters; the latter is essential for the efficiency and convergence of the
diagrammatic Monte Carlo method.

\end{abstract}

\pacs{02.70.Ss, 05.10.Ln}

% 02.70.Ss	Quantum Monte Carlo methods
% 05.10.Ln       Monte Carlo methods

\date{\today}% It is always \today, today,

\maketitle

It has been demonstrated recently that Feynman's diagrammatic
technique can be used for controllable first-principles numeric studies of
lattice fermions (the Hubbard model at $U/t\lesssim 4$) \cite{Hubbard}.
From the diagrammatic point of view, lattice fermions feature a unique
combination of three properties:
(i) Wick's theorem for averages over the non-interacting system
which allows to expand results for the interacting system into
Feynman's diagrammatic series,
(ii) restricted Hilbert space preventing the system from the
Dyson's collapse argument \cite{Dyson52} thereby allowing for the
possibility that the diagrammatic series has a finite convergence radius,
(iii) sign-alternation of diagrams within the same expansion order
leading to the so-called `sign-blessing' phenomenon when high-order diagrams
cancel each other with factorial accuracy.
Spin systems (as well as bosonic ones with restricted filling factor)
share the property (ii), but, speaking generally, not the properties
(i) and (iii). Bosons with unrestricted filling suffer from the Dyson's collapse.
An idea of how the problem of Dyson's collapse could be circumvented for
classical fields and bosons has been proposed recently \cite{PPS};
the price for the finite convergence radius of the regularized theory
is complexity of the interaction vertices which increases with
decreasing the (controllable) systematic error. An outstanding problem
of numerics based on Feynman diagrams, even with
formally convergent series, is the regime of large coupling constants,
where the bare perturbative series fails to converge at small
enough orders of expansion amenable to treatment by the
Diagrammatic Monte Carlo (DiagMC) method \cite{Hubbard}.

In this report, we show how the Popov-Fedotov trick \cite{PF}
of representing spin-1/2 lattice systems with two-component fermions
having imaginary chemical potential and its extensions to other
spin values and projected Hamiltonians \cite{misha},
can be generalized to arbitrary spin, boson and fermion
models with various constraints on the site Fock states.
We would like to stress here the difference in the final formulation
of the problem between the Popov-Fedotov type theories and the slave fermion
approach, see e.g. \cite{slave}. While constraints on the Hilbert space
in the latter method are implemented with the use of additional
local gauge fields, in Popov-Fedotov's (and our) formulation the
same goal is achieved by constructing an equivalent Hamiltonian based
solely on conventional fermionic particles. In the general theory presented
below all constraints on the site Fock states are `relocated'
from the Hilbert space into the system's Hamiltonian by introducing
non-Hermitian terms with imaginary couplings. The power of this approach
is not limited to providing universal tools for constructing the
Feynman's diagrammatic technique with presumably finite convergence radius
for arbitrary spin/boson systems. Yet another crucial feature of the method is
the possibility of absorbing large on-site coupling constants of the original model
into {\it single-particle} energies of additional fictitious fermions. The
remaining interactions between the fermions are either of the order of unity
(in terms of the hopping amplitudes), or scale linearly with temperature.
This feature can prove useful for treating purely fermionic systems---e.g.,
the Hubbard model with large on-site interaction.

{\it Fermionization.} Let us start with bosons and spins.
Here the idea is to represent bosonic/spin degrees of freedom
by internal states of fictitious fermions living on the sites of
the corresponding lattice. To ensure that the physical Hamiltonian
matrix is not modified by the fermionic nature of fictitious fermions
the new Hamiltonian does not have terms involving hopping of particles.
All matrix elements of the original Hamiltonian are then encoded in
interaction processes which change internal degrees of freedom of
dispersionless fermions immediately leading to the
{\it dynamical} equivalence of the original and `fermionized'
Hamiltonians by construction. The problem arises with the
{\it statistical} equivalence. The Hilbert space
of the fermionized Hamiltonian, $H$, contains non-physical Fock states
of the sites (e.g. with occupation numbers other than unity)
and thus its partition function differs from that of the original Hamiltonian.
The trick is to introduce {\it additional} terms, $H_*$, which, on one hand,
produce zero when acting on physical Fock states of the sites
(and thus are dynamically irrelevant for the physical subspace),
while, on the other hand, make the partition function
of $H'=H+H_*$ identical to that of the original Hamiltonian.
The crucial observation is that this goal is readily
achieved with $H_*$ involving only on-site interactions,
 $H_*=\sum_j H_*^{(j)}$  (here $j$ is the site index),
due to the following structure of the trace of $H'$
\begin{equation}
{\rm Tr}\, {\rm e}^{-H'/T} \= Z + \sum_{{\cal N}=1} \sum_{\xi_{\cal N}} \, Z_{\xi_{\cal N}} \, \prod_{j\in \xi_{\cal N}} S_j  \, ,
\label{Tr_H_prime}
\end{equation}
where $Z$ is the physical partition function (sum over all physical states of all the sites), $\xi_{\cal N}$ labels different configurations of ${\cal N}$ non-physical sites,    and $Z_{\xi_{\cal N}}$ is the partition function of a physical system living on a lattice from which all the non-physical sites are removed. Finally,
$S_j$ is a single-site trace over all non-physical states of the site $j$.
The generalization of Popov-Fedotov trick is to chose  $H_*^{(j)}$ in such a
way that $ S_j =0$. We will see that there exists a continuum
of straightforward options for doing that {\it individually} for each
site $j$.

Obviously, the idea of relocating constraints on the site Fock states from the Hilbert space into the Hamiltonian can be used in a more general context. A characteristic example is provided by the $t$-$J$ model dealing with two fermionic components subject to the constraint that there are no doubly occupied sites.
Since there is only {\it one} non-physical Fock state per site, one cannot achieve
the situation of Eq.~(\ref{Tr_H_prime}) with $S_j=0$. The trick is to introduce an auxiliary on-site fermionic mode that does not interact with the rest of the system
and has zero Hamiltonian in the physical sector, i.e. it leads to trivial replacement $Z \to 2^V Z$,
where $V$ is the total number of system sites. For non-physical sites, however,
coupling to the auxiliary mode is chosen in such a way that $S_j=0$. We find it
reasonable to use the term fermionization in this case as well, since the formalism of constructing the Hamiltonian $H'$ here is often the same. Since introducing
an auxiliary mode is not subject to any restrictions, the trick is universal.

{\it Formalism.} Consider first the case involving the above-mentioned auxiliary
fermion mode. One solution is based on Eq.~(\ref{Tr_H_prime}), meaning that
we need the Hamiltonian $H$ to be written in fermionic operators in such a
way that the physical sector is an invariant subspace of $H$.
This is readily achieved by multiplying each elementary term in $H$ by projectors onto the physical states of all the sites this term deals with. Note that for some terms this procedure is redundant; an important example being terms diagonal in the site Fock representation. Then we take
\begin{equation}
H_*^{(j)}\= i\pi T \, P^{(j)}_* \, \tilde{n}_j \, \, ,
\label{tilde_n_term}
\end{equation}
where $ P^{(j)}_*$ is the projector onto non-physical Fock states of the site $j$, and
$ \tilde{n}_j$ is the number operator for the auxiliary fermionic mode.
By construction, $H$ and $H_*^{(j)}$ commute, which brings us to
Eq.~(\ref{Tr_H_prime}). In view of (\ref{tilde_n_term}),
we then have $S_j \propto \sum_{\tilde{n}=0,1} \exp(-i\pi \tilde{n})=0$.

Less restrictively, one can demand that $H$ is only preserving the notion
of a non-physical site, i.e. it does not change the $\xi_{\cal N}$ configuration
but is allowed to change non-physical states of the sites and couple them to
physical ones. Now, in the expression
\begin{equation}
{\rm Tr}\, {\rm e}^{-H'/T} \= 2^{V}Z + \sum_{{\cal N}=1} \sum_{\xi_{\cal N}}
\, 2^{V-N}{\cal Z}_{\xi_{\cal N}} \,  \prod_{j\in \xi_{\cal N}} \tilde{S}_j \, ,
\label{Tr_H_prime}
\end{equation}
${\cal Z}_{\xi_{\cal N}}$ is the  partition of $H$, including non-physical sites, and excluding the auxiliary-fermion sector---the latter is evaluated explicitly due to factorization---for a given set $\xi_{\cal N}$ of non-physical sites;    $\tilde{S}_j = \sum_{\tilde{n}=0,1} \exp(-i\pi \tilde{n})=0$.

Now we turn to a more specific case of spins/bosons
(which allows to avoid the auxiliary fermion trick)
and consider the fermionized Hamiltonian $H$ which includes
both on-site, $H_0$, and inter-site, $H_1$,  parts:
\begin{equation}
H\= H_0+H_1\, ,
\label{H}
\end{equation}
The inter-site Hamiltonian has the following structure
\begin{equation}
H_1\=  \sum_{i\neq j} \, \sum_{\alpha,\beta,\gamma,\delta} \Lambda_{{\alpha,\beta,\gamma,\delta}}^{ij}\, Q _{\alpha\to \beta}^{(i)}
Q _{\gamma\to \delta}^{(j)}  \, .
\label{H_intersite}
\end{equation}
Here $i$ and $j$ are the site labels, Greek letters stand for the {\it physically relevant} fermionic Fock states of a site, and
$\Lambda_{{\alpha,\beta,\gamma,\delta}}^{ij}$ is encoding all
non-local matrix elements of the original physical system.
The on-site operator
$Q _{\alpha\to \beta}^{(i)}$ changes the state of the site $i$ from
$\alpha$ to $\beta$, yielding identical
zero otherwise (most notably, for all the physically irrelevant states of the site).
With the diagrammatic expansion in mind, we need this operator to
be a polynomial in terms of the fermionic creation and annihilation operators.
That can be guaranteed by writing it in the form
\begin{equation}
 Q _{\alpha\to \beta}^{(i)}\= % P_{\beta}^{(i)}
 A _{\alpha\to \beta}^{(i)} P_{\alpha}^{(i)} \, ,
\label{Q}
\end{equation}
where $A _{\alpha\to \beta}^{(i)}$ is the polynomial of the fermionic creation and annihilation operators, such that when acting
on the state $\alpha$ it produces the state $\beta$,
and $P_{\alpha}^{(i)}$ is a projector on the state $\alpha$.

For example, if we want to `fermionize' a system of single-component lattice bosons (with the constraint that the maximal site occupancy is $m_0$),
 the physical state of a site with $m \leq m_0$ bosons is \cite{slave}
(here we suppress the site label for clarity)
 \[
 |\alpha \equiv m \rangle \=  |\, n_m=1;\,  n_l=0, ~l\neq m \, \rangle \, .
 \]
Here $n_l=0,1$ with $0\leq l \leq m_0$ is the site occupation number for the $l$-th fermionic component. Since physically relevant states have exactly one
fermion per site, for single-particle hopping transitions we have
(below $l=m \pm 1$)
\begin{equation}
Q _{m\to l}^{(i)}\= f_{il}^{\dagger} \, f^{\,}_{im} \,
\prod_{ k\neq m,l}\, (1-n_{ik}) \, ,
\label{Q_proj}
\end{equation}
where $f_{im}^{\dagger}$ ($f_{im}$) is the creation (annihilation) operator for the $m$-th fermionic component on the site $i$, and $n_{ik}=f_{ik}^{\dagger}\, f_{ik}$.
[There is no need to include $n_{im}(1-n_{il})$ operators in the state projector
since matrix elements of $Q^{(i)}$ are automatically zero unless $n_{im}=1$ and $n_{il}=0$.]

The on-site Hamiltonian---for simplicity, we discuss here the case
when $H_0$ is diagonal in the site Fock basis---can be cast into
different forms. By insisting that $H$ is zero for non-physical sites
we arrive at
\begin{equation}
H_0\=  \sum_{j,\alpha} \, E_{\alpha}P_{\alpha}^{(j)}\, ,
\label{H_0_1}
\end{equation}
where $E_{\alpha}$ is a real number. As long as all $E_{\alpha}$ are not
large compared to the non-local matrix elements, this form appears to be sufficient.
When some parameters $E_{\alpha}$ become
large (note that for bosons $E_{\alpha}$ describe the on-site interaction),
the form (\ref{H_0_1}) becomes problematic for the diagrammatic technique
because its non-linearity in fermionic occupation numbers entails the
necessity to expand in powers of $E_{\alpha}$.
It turns out that in such cases one can switch to $H_0$ where projectors
are replaced with occupation numbers of fermionic modes, if necessary,
at a  price of adding auxiliary degrees of freedom.
We will start with $H_0$ in the form (\ref{H_0_1}),
and then upgrade the theory to the general case.

Consider the trace of $e^{-H'/T}$ in the site Fock representation. Since all
terms in $H$ contain projectors on the physical states, it gives zero when
acting on any non-physical state of the site. Likewise, by construction all
matrix elements of $H_*$ are zero for each physical site meaning that $[H,\,  H_*]=0$.
As a result, we arrive at the structure of ${\rm Tr}\, {\rm e}^{-H'/T}$
expressed by Eq.~(\ref{Tr_H_prime}). Indeed, for a given configuration
$({\cal N},{\xi_{\cal N}})$ of the non-physical sites, we can ignore
these sites when calculating the trace of $e^{-H/T}$, and ignore physical
sites when calculating the trace of $e^{-H_*/T}$. This leads to the
factorization of the partition function,
$Z(H', \xi_{\cal N})\= Z(H, \xi_{\cal N})\,  Z(H_*, \xi_{\cal N})$,
since traces in $Z(H, \xi_{\cal N})$ and $Z(H_*, \xi_{\cal N}) = \prod_{j\in \xi_{\cal N}} S_j  $ are taken over mutually exclusive sets of sites.

To have $S_j=0$, one can choose $H_*^{(j)}$ in the form
(since $H_*^{(j)}$ acts only on non-physical states it does not need to be
Hermitian)
\begin{equation}
H_*^{(j)}\= -T \sum_{\bar{\alpha}} \, \ln (w_{\bar{\alpha}})\, P_{\bar{\alpha} }^{(j)} \, ,
\label{H_star}
\end{equation}
where $\bar{\alpha}$ labels non-physical states of the site $j$,
and $w_{\bar{\alpha}}$ are certain complex numbers.
With Eq.~(\ref{H_star}) we have
$S_j \= S \= \sum_{\bar{\alpha}} \, w_{\bar{\alpha} }$ and to satisfy the requirement $S=0$
it is necessary and sufficient to choose complex numbers
$w_{\bar{\alpha}}$ in such a way that their sum is zero, $\sum_{\bar{\alpha}} \, w_{\bar{\alpha} }  \= 0$.

One simple possibility is to consider
$w_{\bar{\alpha} } = {\rm e}^{i\phi_{\bar{\alpha} }}$ and tune phases
\begin{equation}
\sum_{\bar{\alpha}} \, {\rm e}^{i\phi_{\bar{\alpha} }}  \= 0 \, .
\label{phases}
\end{equation}
Consider, e.g., the Popov-Fedotov case of spin-1/2 system. The two physical states are
$|1\rangle \, \equiv\,  |n_\uparrow=1,n_\downarrow=0 \rangle $ and $|2\rangle \, \equiv\,  |n_\uparrow=0,n_\downarrow=1 \rangle$.
The two non-physical states are $|3 \rangle \, \equiv\,  |n_\uparrow=1,n_\downarrow=1 \rangle $ and $|4 \rangle \, \equiv\,  |n_\uparrow=0,n_\downarrow=0 \rangle$, so that $P_{3}^{(j)}=n_{j\uparrow} n_{j\downarrow}$ and  $P_{4}^{(j)}=(1-n_{j\uparrow})(1- n_{j\downarrow})$. We see that the original Popov-Fedotov trick corresponds to $\phi_{3}=-\phi_{4}=\pm \pi/2$:
\[
H_*^{(j)}\= \mp \frac{i\pi T}{2}  \left[ P_{3}^{(j)}-P_{4}^{(j)} \right] = \mp \frac{i\pi T}{2}  \left[ n_{j\uparrow} + n_{j\downarrow}  -1 \right] .
\]
Obviously, there is a continuum of other choices. For example, $\phi_{3}=\pi$, $\phi_{4}=0$ yielding $H_*^{(j)}=-iT\pi  n_{j\uparrow} n_{j\downarrow}$.

{\it Eliminating large on-site coupling.} In fermionized spin/boson systems
one can always avoid the problem of large coupling constants associated with
large values of $E_{\alpha}$. Consider lattice bosons with $m_0>1$ where
strong on-site interaction $U$ translates into large values of
$E_{\alpha}$. In this case, we chose $H_0$ in the diagonal bilinear
form of fermionic operators:
\begin{equation}
H_0\=  \sum_{j,m } \, E_{m} n_{j m} \, .
\label{H_0_2}
\end{equation}
Though $H_0$ does not nullify the non-physical states, it still commutes with
the purely local $H_*$ and preserves factorization of the partition function
in Eq.~(\ref{Tr_H_prime}). We simply have to account for $H_0$ energies
when calculating $S_j$-factors. For $H_*^{(j)}$ given by
Eq.~(\ref{H_star}) we then get
\begin{equation}
S\=  \sum_{\bar{\alpha}} \, {\rm e}^{-\eta_{\bar{\alpha}}/T}\, w_{\bar{\alpha} }  \, ,
\label{S_gen}
\end{equation}
\begin{equation}
\eta_{\bar{\alpha}}\= \sum_m \, E_m\langle \bar{\alpha} | n_m | \bar{\alpha}\rangle\, .
\label{sum}
\end{equation}
We see that achieving  $S=0$ is easy if for any state
$| \bar{\alpha}\rangle$ there exists at least one other state
$| \bar{\alpha}' \rangle$ with exactly the same value of $\eta_{\bar{\alpha}}$---states with the same energy can be made canceling each other by using phase factors $w_{\bar{\alpha} } = {\rm e}^{i\phi_{\bar{\alpha} }}$ only.
[Needless to say that the requirement of having two non-physical site states
with exactly the same energy can be always met by simply introducing the
above-discussed auxiliary fermionic mode.]

Less obvious is the fact that there always exists a Popov-Fedotov type solution
for a generic bosonic/spin problem with the Hamiltonian (\ref{H})-(\ref{H_intersite}), (\ref{H_0_2}) and local $H_*^{(j)}$ of the form \cite{Vitya}
\begin{equation}
H_*^{(j)}\= -T \ln (\lambda_j)  ( \hat{n}_j -1)  \, ,\quad  \hat{n}_j\=  \sum_{m } \,
n_{j m}\, .
\label{P_F_gen}
\end{equation}
Indeed, $H_*^{(j)}$ commutes with the physical Hamiltonian and is identically
zero in the physical sector. The expression for $S_j$ then has the form
\begin{equation}
S_j\= {\cal P}_{m_0}(\lambda_j) \, ,
\label{lam}
\end{equation}
where ${\cal P}_{m_0}(\lambda_j)$ is an order-$m_0$ polynomial in $\lambda_j$.
By the central theorem of algebra there exist $m_0$ (including degeneracies)
solutions to (\ref{lam}).
The choice of $\lambda_j$ does not matter; one is free to choose one value
for all sites or work with their arbitrary site distribution.

%For example, this is the case for the spin-1/2 system in the magnetic field.
%Here $m=\,\uparrow,\downarrow$, with $E_{\downarrow}=-E_{\uparrow}$, and
%$\sum_m \, E_m \langle \bar{1} | n_m | \bar{1}\rangle=
%\sum_m \, E_m \langle \bar{2} | n_m | \bar{2}\rangle=0$.

{\it  Second fermionization. Fermionic Hubbard model.} The fermionization technique for bosonic/spin modes can be also used in the context of purely
fermionic systems/modes to eliminate direct on-site interactions
between the fermions. The general principle is to exclude
from the system's Hilbert space states with more than one fermion
on a site by interpreting  them as either hard-core composite bosons
(for an even number of original fermions on a site), or composite fermions
(for larger than unity odd number of original fermions on a site).
The composite nature of the new bosonic/fermionic modes
is not relevant for the final theory which treats bare and composite
particles on equal footing. The hard-core bosonic
modes are then fermionized; we call this procedure {\it second fermionization}.

As an example, consider the fermionic
Hubbard model
\[
H\= -\sum_{<ij>,\sigma} f^{\dagger}_{j\sigma}f_{i\sigma} + \sum_i  \left[ Un_{i\uparrow}n_{i\downarrow}
- \mu_{\uparrow} n_{i\uparrow} - \mu_{\downarrow} n_{i\downarrow}\right] \, ,
\]
where $\sigma=\uparrow, \downarrow$, and  $<ij>$ restricts the summation
to the nearest-neighbor sites (hopping amplitude is set
equal to unity). We are interested in the case when the on-site
interaction $U$ and the chemical potentials $\mu_{\sigma}$ can be large.
We interpret the doubly occupied sites, $n_{i\uparrow}=n_{i\downarrow}=1$,
as containing one hard-core
boson (doublon) with the energy $E^{(b)}=U-\mu_{\uparrow} -\mu_{\downarrow}$. Within this representation, the Hamiltonian is re-written in terms of the fermions subject to the constraint $n_{i\uparrow}+n_{i\downarrow}\leq 1$ and the doublons subject
to the constraint $m_0=1$. The constraints on fermions and doublons---upon fermionization of the latter using fictitious two-component fermions---are implemented by using tools of projectied operators to deal with hopping matrix elements, Eqs.~(\ref{H_intersite}), (\ref{Q}).
The elimination of non-physical sites can then be achieved by the auxiliary fermionic mode. The resulting model thus will involve five, instead of original two, fermionic modes.
However, in the new Hamiltonian all multi-particle coupling constants are equal
to unity (the imaginary coupling constant in $H_*$ scales as $T$).

We are grateful to M. Kiselev and O. Starykh for bringing our attention to the
work by Popov and Fedotov and appreciate useful discussions with V. Gurari and
M. Kiselev.  We acknowledge the hospitality of the Institute for Nuclear Theory, University of Washington and the Aspen Center for Physics.
This work was supported by the National Science Foundation,
grant PHY-1005543, and by a grant from the Army Research Office
with funding from the DARPA OLE program.


\begin{thebibliography}{99}
\bibitem{Vitya}  This fact was realized during a discussion with Victor Gurari.
\end{thebibliography}
\end{document}